\documentclass{article}

\usepackage{microtype}
\usepackage{graphicx}
\usepackage{subfigure}
\usepackage{mathtools}
\usepackage{amsfonts}
\usepackage{booktabs} 

\usepackage{hyperref}

\usepackage[accepted]{icml2020}

\icmltitlerunning{MUSE Microscopy to H\&E Histology Modality Conversion}

\begin{document}

\twocolumn[
\icmltitle{Slide-free MUSE Microscopy to H\&E Histology Modality Conversion \\ via Unpaired Image-to-Image Translation GAN Models}




\begin{icmlauthorlist}
\icmlauthor{Tanishq Abraham}{ucd}
\icmlauthor{Andrew Shaw}{usf,wamri}
\icmlauthor{Daniel O'Connor}{usf}
\icmlauthor{Austin Todd}{utsa}
\icmlauthor{Richard Levenson}{ucdsac}
\end{icmlauthorlist}

\icmlaffiliation{ucd}{Department of Biomedical Engineering, University of California, Davis}
\icmlaffiliation{usf}{University of San Francisco Data Institute}
\icmlaffiliation{wamri}{Wicklow AI in Medicine Research Initiative}
\icmlaffiliation{utsa}{University of Texas, San Antonio Health}
\icmlaffiliation{ucdsac}{Department of Pathology, University of California, Davis}

\icmlcorrespondingauthor{Richard Levenson}{rmlevenson@ucdavis.edu}

\icmlkeywords{CycleGAN, MUSE, Histology, Microcopy, Image-to-Image Translation}

\vskip 0.3in
]



\printAffiliationsAndNotice{}  

\begin{abstract}
MUSE is a novel slide-free imaging technique for histological examination of tissues that can serve as an alternative to traditional histology. In order to bridge the gap between MUSE and traditional histology, we aim to convert MUSE images to resemble authentic hematoxylin- and eosin-stained (H\&E) images. We evaluated four models: a non-machine-learning-based color-mapping unmixing-based tool, CycleGAN, DualGAN, and GANILLA. CycleGAN and GANILLA provided visually compelling results that appropriately transferred H\&E style and preserved MUSE content. Based on training an automated critic on real and generated H\&E images, we determined that CycleGAN demonstrated the best performance. We have also found that MUSE color inversion may be a necessary step for accurate modality conversion to H\&E. We believe that our MUSE-to-H\&E model can help improve adoption of novel slide-free methods by bridging a perceptual gap between MUSE imaging and traditional histology.
\end{abstract}

\section{Introduction}
Microscopy with ultraviolet surface excitation (MUSE) is a novel non-destructive, slide-free tissue imaging modality for histology \cite{fereidouni_microscopy_2017}. Using MUSE instead of conventional histology processing eliminates the need for fixing and thin-sectioning the tissue. While MUSE has been evaluated for many purposes, the current gold standard in medicine and biological research for tissue sample analysis is still based mainly on brightfield imaging of H\&E-stained tissue slides; these dyes color cell nuclei blue and cytoplasm pink, respectively. MUSE dyes, on the other hand, typically involve DAPI or Hoechst dyes for nuclei, and rhodamine for the other tissue components \cite{fereidouni_microscopy_2017}. The resulting images thematically resemble H\&E, but the colors generated by the UV excitation light impinging on these dyes are dramatically different from the traditional brightfield hues.

In order to bridge the gap between MUSE imaging and traditional histological examination, it is possible to digitally modify the MUSE images to match H\&E images. In \cite{fereidouni_microscopy_2017}, a spectral unmixing color mapping model was used, but it required user input of expected colors and is limited to conversion of nuclear and cytoplasm colors, failing to handle cases in which a larger gamut of colors are generated. Therefore, we aim to utilize deep learning methodologies in order to learn the appropriate transformation for generating visually convincing virtual H\&E images that works well on a variety of tissue and cell types.

Deep learning has been used successfully in microscopy modality conversion tasks like the one we present here, \cite{rivenson_phasestain_2019,borhani_digital_2019,rivenson_virtual_2019}. These modality conversion algorithms often use a generative adversarial network (GAN) framework \cite{goodfellow_generative_2014}. In this case, the generator needs to be trained with the input modality image and a corresponding output modality image. Therefore, paired image datasets are required for modality-converting GANs. Unfortunately,  it is not possible to obtain exact pixel-aligned H\&E and MUSE images. Therefore, we investigated unpaired image-to-image translation methods that eliminate the need for precisely paired datasets.

We propose a framework for training and applying an image-to-image translation GAN-based algorithm for successful conversion of MUSE images to virtual H\&E images. We evaluated CycleGAN \cite{zhu_unpaired_2017}, DualGAN \cite{yi_dualgan_2017}, and GANILLA \cite{hicsonmez_ganilla_2020} for MUSE-to-H\&E conversion. We hope that our framework will help catalyze the adoption of MUSE and improve the efficiency of the pathologist’s workflow.

\section{Methodology}
We define two image domains, one for MUSE images ($X$), and one for H\&E images ($Y$). We attempt to determine the transformation $G\colon X\rightarrow Y$. In our framework, we have two tasks. One task is to learn a generator $G_X\colon X \rightarrow Y$ that maps $x\in X$ to $y\in Y$. The auxiliary task is to learn a generator $G_Y \colon Y \rightarrow X$. Additionally, we have the adversarial discriminators $D_X$ and $D_Y$. $D_X$ discriminates between the fake outputs of $G_X$ and real images from domain $Y$. Conversely, $D_Y$ discriminates between the fake outputs of $G_Y$ and real images from domain $X$. These two GANs form the training framework for MUSE-to-H\&E conversion. CycleGAN, DualGAN, and GANILLA all follow this framework and only differ slightly in model architectures, loss functions, and training procedures.

\subsection{CycleGAN}
\label{sectioncyclegan}
CycleGAN exploits the cycle-consistency property that $G_Y (G_X (x))\approx x$ and $G_X (G_Y (y)) \approx y$. This constraint can be expressed as the following loss:

$$\begin{aligned} \mathcal{L}_{\text{cycle}}\left(G_{X}, G_{Y}\right)=& \mathbb{E}_{x \sim p_{\text {data}}(x)}\left[\left\|G_{Y}\left(G_{X}(x)\right)-x\right\|_{1}\right] \\ &+\mathbb{E}_{y \sim p_{\text {data}}(y)}\left[\left\|G_{X}\left(G_{Y}(y)\right)-y\right\|_{1}\right] \end{aligned}$$

where $\|\cdot\|_{1}$ is the $L_{1}$ norm.
Additionally, the GANs are trained with the traditional adversarial losses \cite{zhu_unpaired_2017}. Finally, for regularization, we impose an “identity” constraint:

$$\begin{aligned}
\mathcal{L}_{\text{identity}}\left(G_{X}, G_{Y}\right) &=\mathbb{E}_{y \sim p_{\text {data}}(y)}\left[\left\|G_{X}(y)-y\right\|_{1}\right] \\
&+\mathbb{E}_{x \sim p_{\text {data}}(x)}\left[\left\|G_{Y}(x)-x\right\|_{1}\right]
\end{aligned}$$

The generator architecture is a ResNet-based fully convolutional network described in \cite{zhu_unpaired_2017}. A 70x70 PatchGAN \cite{isola_image--image_2017} is used for the discriminator. The same loss function and optimizer as described in the original paper \cite{zhu_unpaired_2017} was used. The learning rate was fixed at 2e-4 the first 100 epochs and linearly decayed to zero in the next 100 epochs, like \cite{zhu_unpaired_2017}.

\subsection{DualGAN}
DualGAN \cite{yi_dualgan_2017} also solves the same task as CycleGAN, while using Wasserstein GANs \cite{arjovsky_wasserstein_2017}. DualGAN also uses a reconstruction loss, which is similar to CycleGAN’s cycle-consistency loss. The generator architecture is a U-net \cite{ronneberger_u-net_2015}, and the discriminator is a 70x70 PatchGAN \cite{isola_image--image_2017}. The model was trained with the Adam optimizer for 200 epochs similar to the CycleGAN (\hyperref[sectioncyclegan]{Section 2.1}).

\subsection{GANILLA}
GANILLA \cite{hicsonmez_ganilla_2020} is a variant architecture for the CycleGAN model designed to appropriately transfer the content to the stylized image. See \cite{hicsonmez_ganilla_2020} for generator architecture details. The discriminator is a 70x70 PatchGAN. The model was trained for 200 epochs in an identical manner as the CycleGAN  (\hyperref[sectioncyclegan]{Section 2.1}).

\subsection{Color mapping tool}
As a baseline, we used a color mapping tool using spectral unmixing algorithms, as previously described in \cite{fereidouni_microscopy_2017}.

\subsection{Tiled inference}
\label{inference}
We performed GAN model inference on overlapping tiles with stride 256. The generator was applied to each patch, yielding a 19x19 array of overlapping predicted H\&E patches. A 5120x5120 generated H\&E montage was then constructed, with each pixel intensity value in the montage being a weighted average of intensity values from the H\&E patches which overlapped at the given pixel location. The pixel intensity values from each contributing patch were weighted proportionally to $ \exp({{-d^2}/{2\sigma^2}})$ where $d$ is the distance from the given pixel location to the center of the contributing patch. The weights in the weighted average were normalized to sum to 1. The parameter $\sigma$ was set to 128 pixels.

\subsection{Quantitative evaluation of the models}
An external critic model (70x70 PatchGAN with a fully connected layer) was trained to quantitatively evaluate how “real" the outputs of the various models look. We used accuracy and a binary cross-entropy loss from the critic as quantitative measures to compare the quality of the generative models. We trained a separate critic on the predictions for each model to keep results independent. Each critic were trained for 20 epochs with a 0.001 learning rate (one-cycle learning rate schedule). Each dataset consisted of “fake" H\&E images generated from the test set and real H\&E images from the train set. It was a balanced dataset with an 80/20 dataset split.

\begin{figure}[hb]
\centerline{\includegraphics[width=\columnwidth]{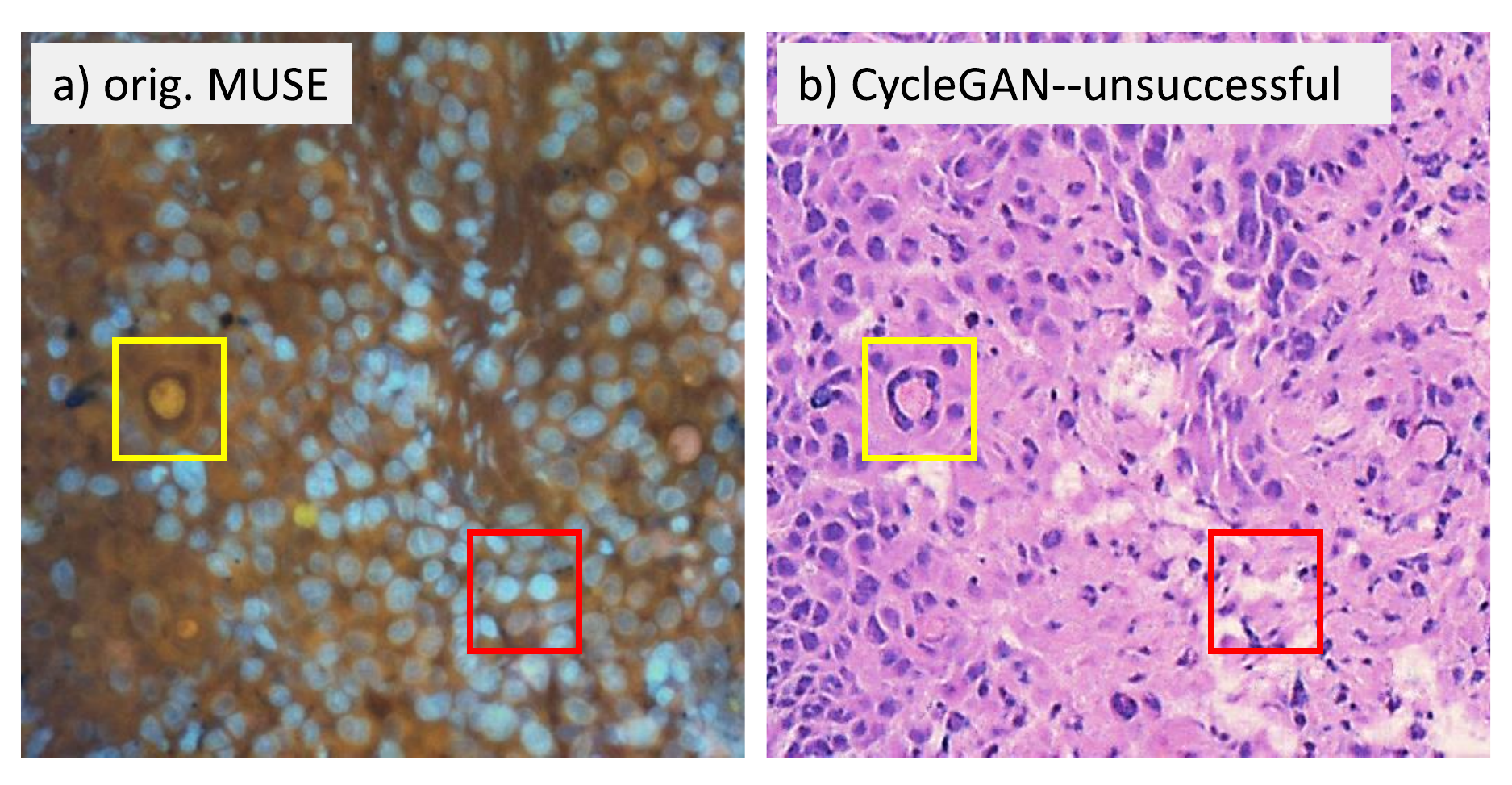}}
\caption{Training CycleGAN on unprocessed MUSE images}
\label{Figure1}
\end{figure}

\subsection{Datasets and implementation}
\label{dataset}
The H\&E data came from a region in a single whole-slide image of human kidney with urothelial cell carcinoma. The MUSE data came from a single surface image of similar tissue. We obtained 512x512 tiles from the images, resulting in 344 H\&E tiles and 136 MUSE tiles. The tiles were randomly cropped into 256x256 images when loaded into the model. Code was implemented with PyTorch 1.4.0 \cite{paszke_pytorch_2019}, and fastai \cite{howard_fastai_2020}.

\section{Results}

\subsection{Training on unprocessed MUSE images}
In \hyperref[Figure1]{Figure 1}, results on the test dataset after training a CycleGAN on MUSE and H\&E images are shown. With close inspection, it is evident that the generated H\&E images do not appropriately transfer the content of the original MUSE image. Bright in-focus nuclei are converted to white spaces in the virtual H\&E image (boxed in red). On the other hand, the darker regions are converted to nuclei in the H\&E image (boxed in yellow). The overall trend that the CycleGAN followed was converting brighter regions to background white spaces, and darker regions to nuclei. We have observed that using color- and intensity-inverted MUSE images greatly improves training and subsequent models were trained on inverted MUSE images.

\subsection{MUSE-to-H\&E translation}
\label{translation}
We trained a CycleGAN, DualGAN, and GANILLA model on the MUSE and H\&E image dataset (\hyperref[dataset]{Section 2.7}), and performed inference on the test dataset, which is a 5120x5120 image. Individual 512x512 tiles were inputted into the model.

\begin{figure}[t]
\centerline{\includegraphics[width=\columnwidth]{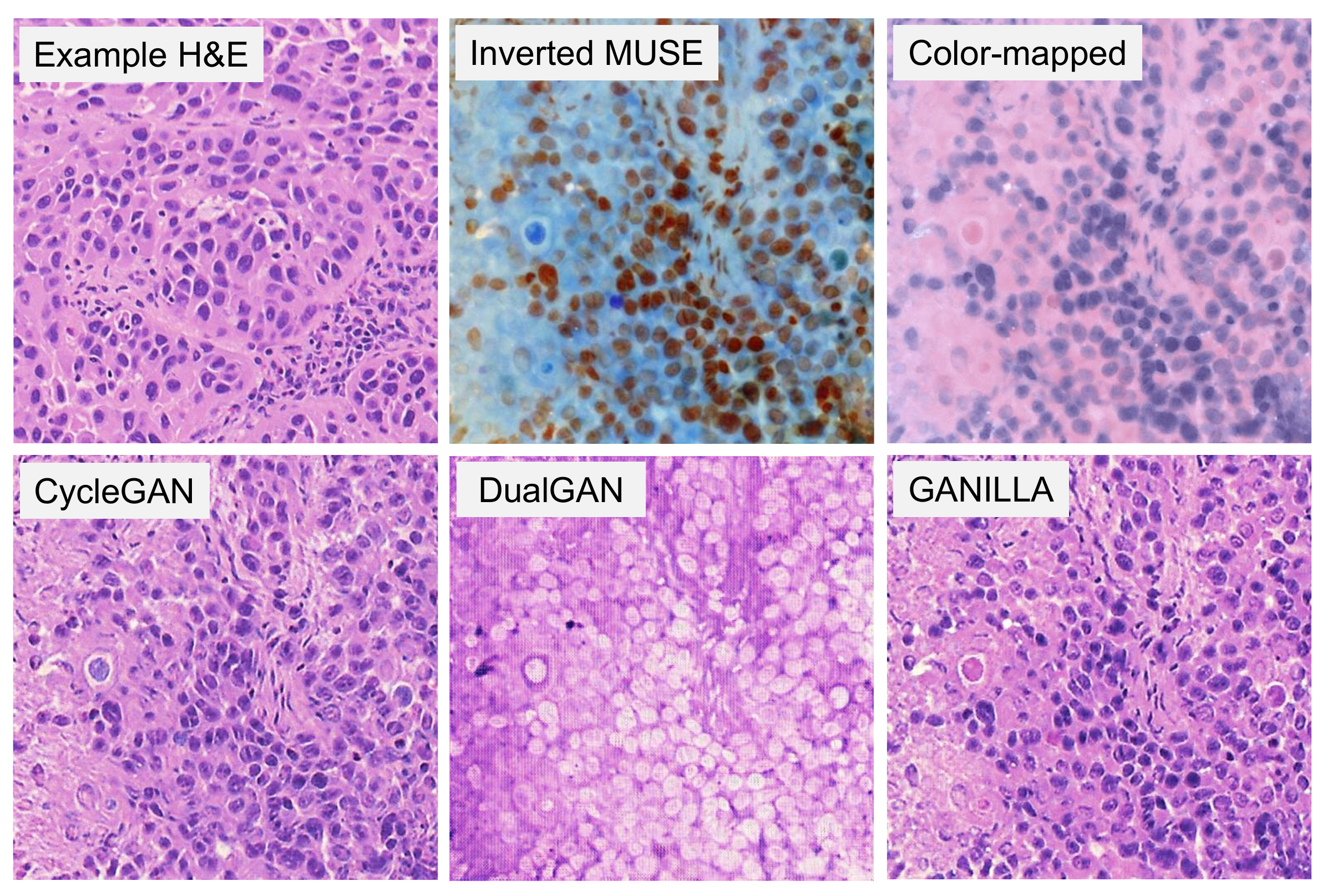}}
\caption{Comparison of MUSE-to-H\&E translation models}
\label{Figure2}
\end{figure}

In \hyperref[Figure2]{Figure 2}, we can see that the CycleGAN and GANILLA models provided visually compelling results that appropriately transfer style and content. The model successfully converted MUSE representations of cancer cells, inflammatory cells, and connective tissue to the corresponding H\&E representations. However, DualGANs performed poorly, with weak transfer of style, and many artifacts. Finally, CycleGAN and GANILLA performed better than the traditional color-mapping baseline.

\subsection{Inference}
\begin{figure}[b]
\centerline{\includegraphics[width=\columnwidth]{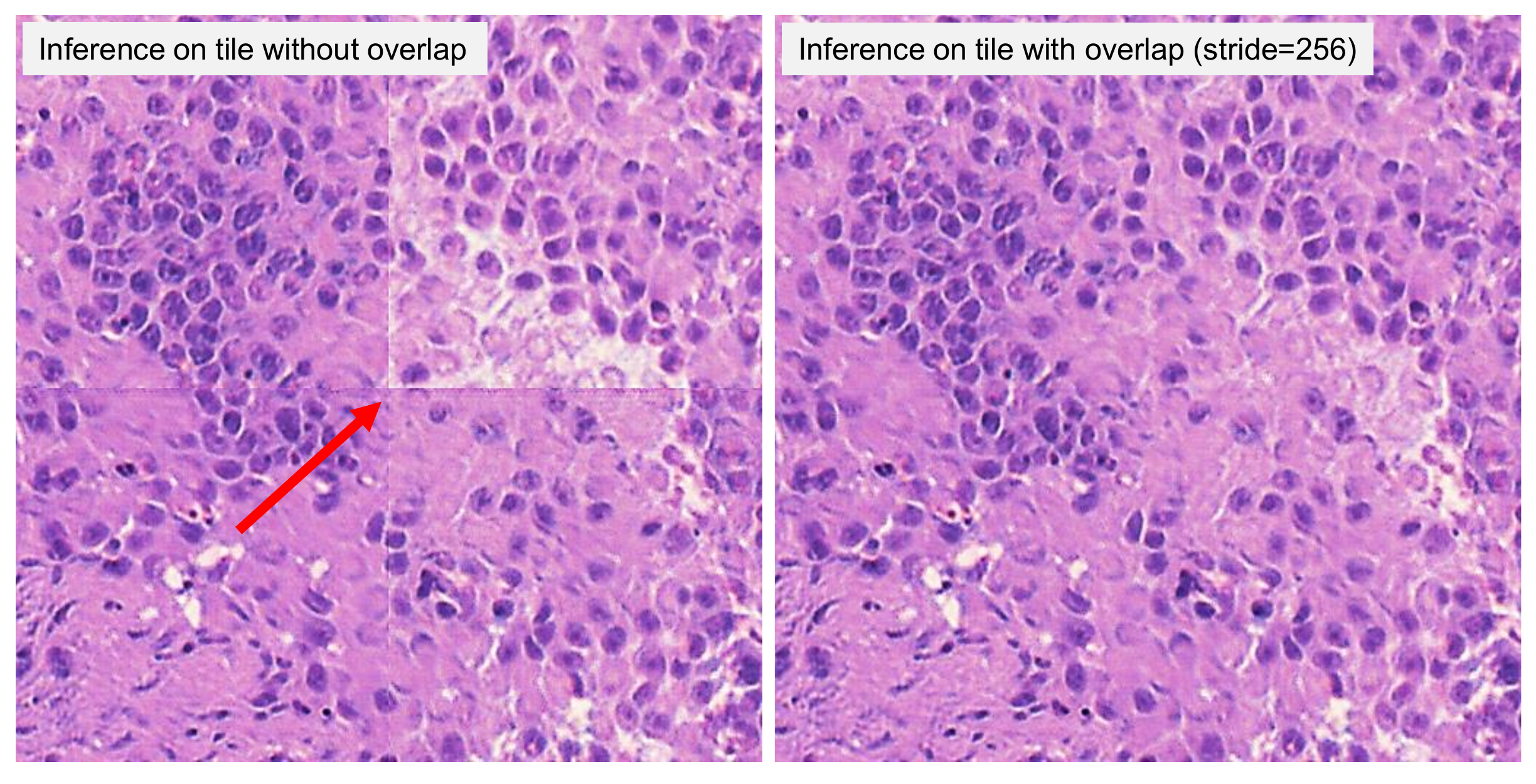}}
\caption{Demonstration of tiling artifacts}
\label{Figure3}
\end{figure}

We have tested inference with a single 5120x5120 image. As the generators are fully convolutional networks, variable sizes are allowed for these models (though the scale must remain same). However, the full region cannot be inputted into the model due to memory constraints. While the models performed well on individual 512x512 patches, we observed (\hyperref[Figure3]{Figure 3}) that the montage had artifacts near the edges of the individual patches (tiling artifacts), and the predictions are inconsistent in color and style between tiles.

\begin{figure}[t]
\centerline{\includegraphics[width=\columnwidth]{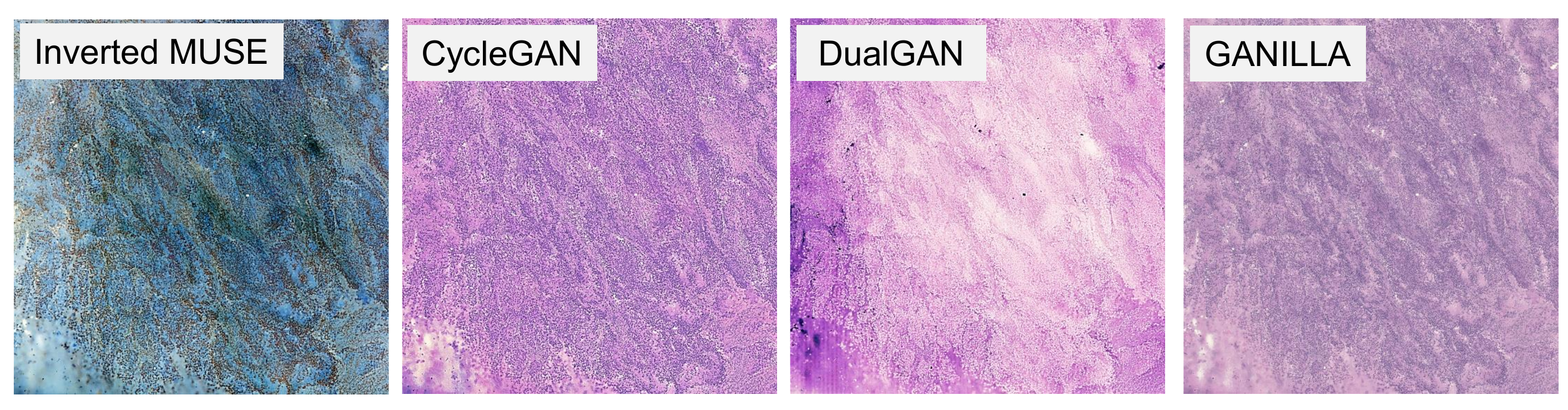}}
\caption{Montages generated from predictions on overlapping 512x512 tiles with stride 256.}
\label{Figure4}
\end{figure}

In order to resolve these problems, we performed model inference on overlapping tiles with stride 256 as explained in \hyperref[inference]{Section 2.5}. \hyperref[Figure3]{Figure 3} demonstrates how this blending approach suppressed the emergence of tiling artifacts. \hyperref[Figure4]{Figure 4} shows that the final generated montages were much more consistent in style and color throughout the montage.

\subsection{Critic training}

Using the H\&E generated results of the CycleGAN, GANILLA and DualGAN models, we trained three separate external critics to objectively measure the quality of the generated images. A fourth critic was trained on images from the color mapping tool as a baseline comparison. In this experiment, we would expect the critic model to fail more often if the model outputs are higher quality, that is, resemble H\&E images more closely.

\begin{figure}[b]
\centerline{\includegraphics[width=\columnwidth]{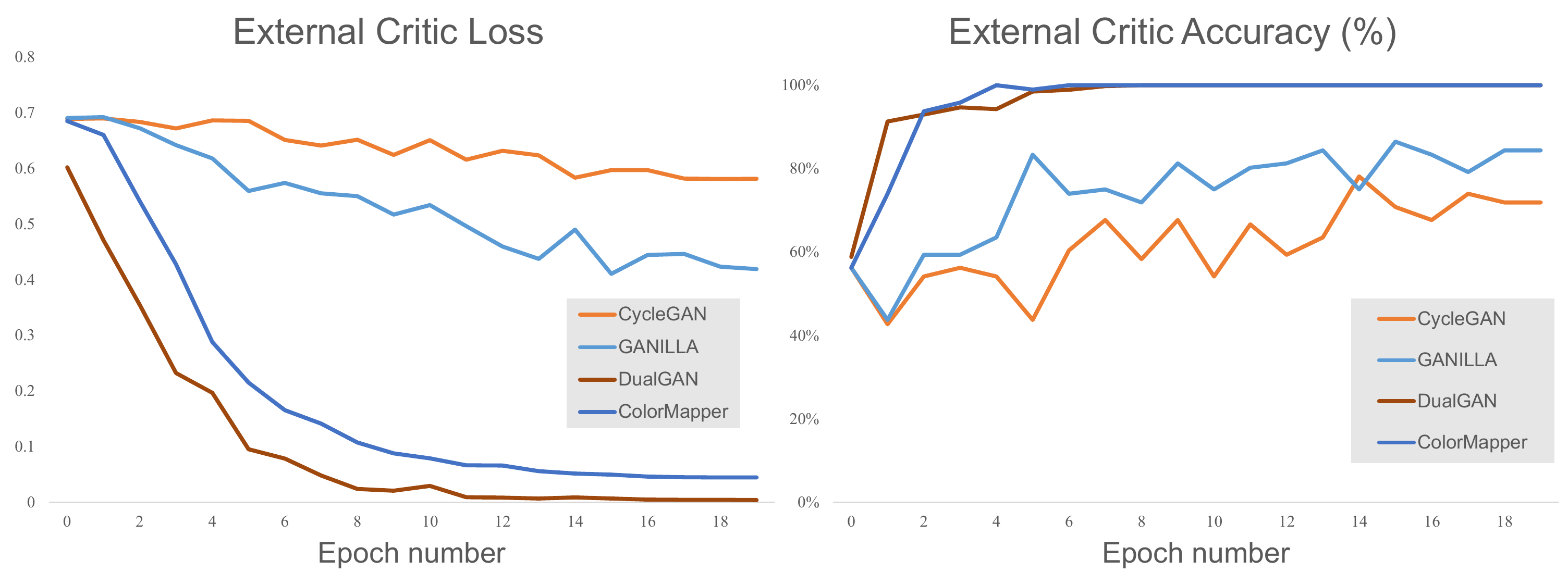}}
\caption{Quantitative evaluation of the models via real/fake H\&E critic training}
\label{Figure5}
\end{figure}

\hyperref[Figure5]{Figure 5} shows the graphs of the validation loss and accuracy while \hyperref[Table1]{Table 1} presents the accuracy from critic training. They show that the critic performed more poorly on CycleGAN and GANILLA images. The DualGAN was not able to fool the critic because of its poor performance in producing a convincing color conversion. Interestingly, DualGAN performed worse than the color mapper baseline. Overall, the critic had the hardest time identifying the CycleGAN model as “fake”, which seems to suggest this model produced the most realistic images. These results supports the conclusions from qualitative analysis in \hyperref[translation]{Section 3.2}.

\section{Discussion}

In this study, MUSE modality conversion using unpaired image-to-image translation techniques was performed in order to generate virtual H\&E images. We qualitatively observed that the GAN-based models studied here produce visually compelling results, with CycleGAN providing the best results.

For proper training and inference of the models tested here, inverting the MUSE images was required. This is likely because the CycleGAN cannot learn the association between brighter nuclei in MUSE to darker nuclei in H\&E. It assumed all bright objects in MUSE must be background in H\&E, while dark background objects in MUSE must be tissue in H\&E. We found this content preservation problem especially prevalent in DualGANs. In future work, additional constraints, such as the saliency constraint introduced in \cite{li_unsupervised_2019}, may be tested in order to directly convert unprocessed MUSE images to virtual H\&E images.

\vspace{-1em}
\begin{table}[t] \centering
\caption{Negative Critic Accuracy (\%)}
\label{Table1}
\begin{tabular}{l l l l l}
\toprule
& \multicolumn{4}{c}{Epoch \#} \\
\cmidrule{2-5} 
GAN Loss & 1 & 5 & 10 & 20 \\
\midrule 
CycleGAN & 56.3 & 54.1 & 67.7 & 71.9 \\
GANILLA & 56.3 & 63.5 & 81.2 & 84.4 \\
DualGAN & 58.1 & 94.2 & 100 & 100 \\
Color Mapper & 56.3 & 100 & 100 & 100 \\
\end{tabular}
\end{table}
\vspace{1em}

A major challenge with training unpaired image-to-image translation models is the lack of quantitative metrics. Most approaches for quantifying model performance relied on crowdsourcing approaches (e.g. Amazon Mechanical Turk) to rate quality of the produced images. However, with difficult-to-interpret histological images, this is not an option. Most microscopy modality conversion studies \cite{rivenson_phasestain_2019,borhani_digital_2019,rivenson_virtual_2019} had paired data and therefore quantitatively evaluated via structural and perceptual similarity metrics like SSIM and PSNR \cite{zhou_wang_image_2004}. However, there are some key structural differences between MUSE and H\&E images. This would mean that visually compelling virtual H\&E images that also preserve structural content may not have high perceptual similarity scores. Instead, we relied on an independently trained critic model to estimate image quality and perceptual similarity. While we found the results to be very consistent with our visual inspection, it is important to note that it is not a perfect metric and does not account for GAN “hallucinations" or preservation of content. The best metric is still visual inspection by human beings. Future work will quantitatively evaluate image-to-image translation models with pathologist ratings and interpretation.

Another key consideration during the development of these models is model inference. We expect users to be able to select regions of interest from a whole image to convert to virtual H\&E almost instantly. Currently, this is still a challenge that needs to be addressed (CycleGAN on 5120x5120 with stride 512 took 12.7 s on NVIDIA TITAN RTX). Future work will analyze how model inference can be sped up while minimizing the trade-off regarding montage consistency.

\section{Conclusion}
In conclusion, we have tested three unpaired image-to-image translation models for MUSE-to-H\&E conversion. We observed that color- and intensity-inversion is an essential preprocessing step when training with MUSE images. Additionally, we used a semi-quantitative method of evaluating the models and determined CycleGANs obtain the best results. We hope our framework can help improve the efficiency of the pathologist workflow by bridging the gap between MUSE and traditional histology.

\clearpage

\bibliography{bib}
\bibliographystyle{icml2020}

\end{document}